\documentclass{article}

\usepackage{arxiv}

\usepackage[utf8]{inputenc} 
\usepackage[T1]{fontenc}    
\usepackage{hyperref}       
\usepackage{url}            
\usepackage{booktabs}       
\usepackage{amsfonts}       
\usepackage{nicefrac}       
\usepackage{microtype}      
\usepackage{lipsum}		
\usepackage{graphicx}
\usepackage[numbers]{natbib}
\usepackage{doi}

\usepackage{float}
\usepackage{tabularx}
\usepackage{amsmath}
\usepackage{blindtext}

\title{An analysis of indifference curves and areas from a human nutrition perspective}

\author{ \href{https://orcid.org/0000-0002-0580-4871}{\includegraphics[scale=0.06]{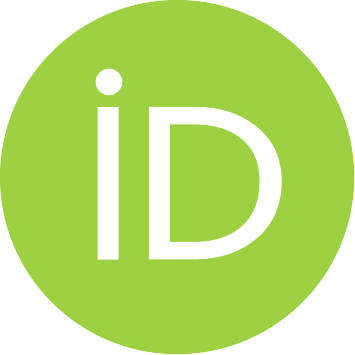}
\hspace{1mm}Diego~Roldán}\thanks{Use footnote for providing further information about author (webpage, alternative
		address)---\emph{not} for acknowledging funding agencies.} \\
	Department of Regional Economy Research\\
	University of Cuenca\\
	\texttt{diego.roldan@ucuenca.edu.ec} \\
	\And
	\href{https://orcid.org/0000-0001-5466-9998}{\includegraphics[scale=0.06]{orcid.pdf}\hspace{1mm}Angélica~Abad Cisneros} \\
	Department of Regional Economy Research\\
	\texttt{angelica.abad@ucuenca.edu.ec} \\
	\And
	\href{https://orcid.org/0000-0001-5180-7363}{\includegraphics[scale=0.06]{orcid.pdf}\hspace{1mm}Francisco~Roldán-Aráuz} \\
	Department of Regional Economy Research\\
	\texttt{francisco.roldan.abad@ucuenca.edu.ec} \\
	\And
	\href{https://orcid.org/0000-0003-3489-9838}{\includegraphics[scale=0.06]{orcid.pdf}\hspace{1mm}Samantha~Leta Angamarca} \\
	Department of Regional Economy Research\\
	\texttt{samantha.leta24@gmail.com} \\
	\And
	\href{https://orcid.org/0000-0002-5846-8744}{\includegraphics[scale=0.06]{orcid.pdf}\hspace{1mm}Anahí~Ramírez Zambrano} \\
	Department of Regional Economy Research\\
	\texttt{aniramirez62@yahoo.com} \\
}

\renewcommand{\shorttitle}{Indifference curves and areas}

\hypersetup{
pdftitle={A template for the arxiv style},
pdfsubject={q-bio.NC, q-bio.QM},
pdfauthor={David S.~Hippocampus, Elias D.~Striatum},
pdfkeywords={First keyword, Second keyword, More},
}

\begin{document}
\maketitle

\begin{abstract}
Through documentary research and interviews with nutrition experts, we found that all nutrients have two thresholds, the Recommended Daily Allowance (RDA) and the Tolerable Upper Intake Level (UL). Intake less than the RDA or more than the UL negatively affects health. Intake quantities of nutrients within these limits covers 100\% of the objective physiological needs without negative repercussions. These characteristics, and others, are common knowledge among nutrition experts; however, these are not adequately reflected in the microeconomics models that study these needs. We conclude that the generalized presence of these thresholds determines the existence of significant \textit{indifference areas} that should be added to the microeconomics models of the \textit{indifference curves}, thus improving the modelling of reality.
\end{abstract}



\section{Introduction}
In microeconomics, indifference curves are models that allow us to analyse the characteristics of utility levels obtained from the combined consumption of commodities. According to conventional theory, combinations of commodities that generate the same level of utility give rise to indifference curves. However, the empirical evidence points to the existence of what in this study we call \textit{indifference areas}, with important implications, particularly in the economic modelling of human nutrition. The explanation of these areas is the focus of this article.

A commodity's utility comes from the level of satisfaction of a human need, understood as a concrete or abstract element that must be present to achieve well-being \cite{ref01}. Towards the middle of the 20th century, Maslow \cite{ref02} classified human needs into five hierarchical levels: physiological needs, safety, love, esteem, and self-actualization. Later, Maslow \cite{ref03} added one more step, proposing self-transcendence needs as a category above the need for self-actualization \cite{ref04}.

According to Maslow, fundamental needs are innate and are the same across ages and cultures \cite{ref05}, a claim that was rigorously tested by Tay and Diener \cite{ref06} with a sample that included individuals from 123 countries. On this point, Gough \cite{ref01} maintains that needs are universal and differ from specific need-satisfiers, which are variable and local; this coincides with the work of Max-Neef \cite{ref07}, who points out that "fundamental human needs are finite, few and classifiable. [They] are the same in all cultures and all historical periods. What changes, both over time and through cultures, is the way or the mean (need satisfiers) by which the needs are satisfied" (p. 19). The current study focuses on physiological needs, and more specifically, on nutritional needs, since these are "the most prepotent of all needs" \cite{ref02}.

Historically, in order to satisfy physiological needs (demand), humanity turned the supply of goods and services from nature \cite{ref08}. Nowadays, technological developments in various disciplines such as agriculture, biology, medicine, and economics allow us to meet some of these needs through artificial means. However, technological development, the significant increase in global population, and the growth of per capita consumption place critical pressure on natural resources, to the extent that our ecological footprint endangers our own food security \cite{ref09}.

Modernity has involved the industrialization of food, which does not necessarily imply a healthy diet. Over the last few decades, the increasing presence of processed and ultra-processed foods in the market has been notable \cite{ref10,ref11}. Some of these patterns are conducive to poor eating habits, leading to diseases that have a significant impact on society and are a concern for governments around the planet; therefore, most countries try to promote healthy eating through the dissemination of information, such as through mandatory nutritional descriptions on the labels of certain commercialized foods \cite{ref12}, which are common on prepackaged food for infants, to cite just one example.

Poor eating habits are often the result of people resorting to food preferences that are palatable (desire) but do not coincide with objective nutritional needs, although unhealthy diets are also known to be a feature of poverty. The negative consequences of an unhealthy diet include obesity, high cholesterol, heart problems, diabetes, hypertension, dyslipidemia, anaemia, infections, cancer, etc. According to the World Health Organization \cite{ref13}, “1.9 billion adults are overweight or obese, while 462 million are underweight. […] Globally in 2020, 149 million children under 5 were estimated to be stunted (too short for age), 45 million were estimated to be wasted (too thin for height), and 38.9 million were overweight or obese.”

A healthy diet implies abandoning subjective desire and opting to satisfy the objective need of one’s body. Titchenal \cite{ref14} argues that “the appropriate amount of protein in a person’s diet is that which maintains a balance between what is taken in and what is used.” In short, when dietary desires (psychological impulse) do not coincide with (nutritional) needs, health problems that can be very serious may develop. Although a large share of the population does not act responsibly in terms of diet, there are many people who consciously strive for a healthy diet, not only for themselves but also for the children, sick, and elderly in their care. They are the population of our research.

It is important to differentiate between food and nutrients. Food refers to the items we eat or drink, such as bread, fruit, vegetables, meat, or milk; nutrients are the components of food that are metabolized and contribute to bodily processes on the cellular level \cite{ref15}. There are two types of nutrients: macronutrients and micronutrients. Macronutrients include proteins, carbohydrates, lipids, and water. Micronutrients include vitamins, and minerals \cite{ref16, ref17}. A good diet involves ingesting the necessary daily amount of each of these nutrients and avoiding deficiencies or excesses that may lead to disease \cite{ref18}. The body absorbs various nutrients from food through digestion or directly from dietary supplements.

\subsection{Indifference curves}

Stiglitz and Walsh \cite{ref19} describe utility as the benefit of consuming commodities in combination with one another. There is no direct way to measure the utility generated, although an indirect way is through people’s \textit{willingness to pay}. According to the \textit{law of diminishing marginal utility}, as the consumption of a bundle of commodities increases, the utility contributed by each unit decreases.

In microeconomics, it is argued that there are several combinations, or bundles, of goods and services that provide a consumer with equal levels of utility \cite{ref20} —in other words, need-satisfiers. A \textit{consumption set X} is a consumption bundle or a subset of the commodity space $\mathbb{R}^L$ that a consumer selects (prefers) and uses.

The aim in microeconomics is to model consumer behaviour under the assumption of rationality, which states that a consumer seeks to maximize the level of utility obtained from the different possible consumption sets $X$, subject to restrictions such as budget constraints. Mas-Colell et al. \cite{ref20} put forward other examples of restrictions, such as the fact that it is not possible to rest more than 24 hours each day and that one’s consumption level must be nonnegative.

In this context, it is possible to assume that different consumption bundles will provide the same level of utility (satisfaction) to the consumer (an \textit{indifference set}). If we consider a bundle of two commodities ($x_1, x_2$), the set of all combinations that provide the same level of utility $U$ \cite{ref21, ref22} are the bundles for which the consumer is indifferent, and these form an indifference curve \cite{ref23}. “One can think of indifference curves as being level sets of the utility function” \cite{ref24}. These curves are very useful for analysing consumer behaviour.

It is common for microeconomics textbooks to begin by introducing consumer preference relations and some of the \textit{basic properties} that are assumed to model this behaviour, that is, what are or should be the most relevant characteristics. For Mas-Colell et al. \cite{ref20}, two properties need to be discussed in relation to this: “monotonicity (or its weaker version, local non-satiation) and convexity.” Varian \cite{ref23} states that “what other properties do indifference curves have? In the abstract, the answer is: not many.”

\subsection{Principle of insatiability}

The insatiability principle (also called non-satiation; \cite{ref20}) holds that “the more, the better” \cite{ref19}. This principle implies that the greater the distance of an indifference curve from the origin, the higher the level of utility. Mas-Colell and colleagues clarify that “The assumption that preferences are monotone is satisfied as long as commodities are "goods" rather than "bads".” For example, if we refer to garbage, it should be analysed as “no garbage”. Both Varian \cite{ref24} and Mas-Colell et al. \cite{ref20} specify this principle in an analysis of \textit{local non-satiation}, “and, hence, of monotonicity”, and deduce that it excludes “thick” indifference sets. According to Varian \cite{ref23}, an important principle is that indifference curves cannot cross each other because they represent distinct levels of preference, as can be inferred from the principle of monotonicity. 

\subsection{Convexity}

\textit{Decreasing marginal utility} implies that each additional unit of a good consumed provides less utility or satisfaction. In other words, when the satisfaction obtained from substituting one good for another remains unchanged, the satisfaction will depend on the previous levels of utility produced by each good. According to Mas-Colell et al. \cite{ref20}, the convexity assumption is concerned with consumer trade-offs when choosing different commodities. This assumption is a central hypothesis in economics that can be interpreted in terms of \textit{diminishing marginal rates of substitution}, i.e. in the case of two commodities, “from any initial consumption situation x, it requires increasingly larger amounts of one commodity to compensate for successive unit losses of the other.” However, there are studies that show substitutions that do not comply with this behaviour. For example, Miljkovic \cite{ref25} analyses the case of food-addicted consumers whose compulsive behaviour violates this assumption of convexity.

\subsection{Cases of interest}

It is common for microeconomics textbooks to explain some special cases. According to Varian \cite{ref23}, there are particular relations of goods that “are \textit{perfect substitutes} if the consumer is willing to substitute one good for the other at a \textit{constant rate}.” In this case, the relationship has a constant slope. In another case, two commodities are \textit{perfect complements} when they are always consumed together in fixed proportions. A common example is shoes: right and left shoes are generally useful only if they come in pairs, and having more left shoes does not increase the utility of the good. For this type of commodity, the indifference curves are L-shaped, with the vertex (total utility) of the “L” occurring where the number of left and right shoes is equal. In this case, the utility is equal only to the number of complete pairs.

Another exception is \textit{neutral goods}, whose consumption does not affect overall satisfaction. “A commodity x1 is a neutral good if the consumer does not care about it one way or the other” \cite{ref23}. If we combine a neutral good $x_1$ (e.g. anchovies) with a non-neutral good $x_2$ (e.g. pepperoni), the “indifference curves will be vertical lines” to the axis of the non-neutral good $x_2$. That is, the amount of $x_1$ (anchovies) does not affect the total utility.

Of the different groups of needs identified by Maslow, physiological needs constitute the foundation of well-being. When this level is reasonably satisfied, the other levels take on importance. Nutrition is part of this group of fundamental needs. Hence, it is worth questioning whether the usual economic modelling of indifference curves in microeconomics textbooks is helpful to explain the behavior of consumers who base their rationality and dietary preferences on the satisfaction of nutritional needs. Our hypothesis is that traditional models fail to do it adequately since they ignore important characteristics present in the actual consumption of nutrients.

To test this hypothesis, we use objective information on the nutritional requirements of the human body gathered through an analysis of the specialized literature and interviews with experts. It is important to point out that the analysis is carried out exclusively from the perspective of nutrient needs, i.e., of the components of food rather than the whole (An analysis focused on components is not new in economic studies. Heckscher (1933) and Ohlin (1919) did it by analysing international trade from the perspective of production factors and not of the products themselves \cite{ref35}. This perspective allowed a better understanding of international trade compared to the comparative advantages proposed by Ricardo.).  Studying dietary needs from the perspective of nutrients allows a better understanding of the process of meeting nutritional needs. In the three sections that follow, we first explain how the information collected was analysed, then we present the results of the research and, finally, discuss the findings in light of the literature and present our conclusions.

\section{Materials and Methods}

To establish the optimal intake values for nutrients to maintain a healthy human body, an analysis of the specialized literature on food, metabolism, and nutrition was carried out, in addition to nine semi-structured interviews with experts. We consulted the following sources during the documentary research: \textit{Human Nutrition: 2020 Edition} by Titchenal \cite{ref14}, \textit{Krause's Food \& The Nutrition Care Process (14th ed.)} by Mahan and Raymond \cite{ref26}, \textit{Alimentacion And Dietotherapy} (4th ed.) by Cervera et al. \cite{ref27}. Digital information from the Mayo Clinic \cite{ref28} as well as Healthline \cite{ref29} was also reviewed. This allowed us to gain an understanding of human nutritional needs, the ways of satisfying these, and the repercussions of inadequate or excessive intake. 

The documentary research allowed us to identify essential nutrients, their characteristics, typology, as well as the possibility of substitutions between them. Their physiological benefits (objectives) were also identified, i.e., the usefulness of each one, the amounts required, and repercussions due to under- or over-intake, allowing us to sketch a general model of the utility of these in terms of human nutrition.

Once the documentary analysis was completed, we conducted nine interviews with experts, namely university professors with professional experience in the fields of metabolism and nutrition. The interviews sought to validate and refine the model constructed from the documentary analysis. The interviewees were specifically asked about the adequacy of the proposed functional forms (abstract spatial representation) used to model nutritional utility, as well as to collect additional information that would allow us to test our hypothesis. As is common in qualitative methodologies, after each interview we further refined our conclusions in accordance with the saturation principle (In qualitative research, the saturation principle proposes that data collection can be concluded when the researchers stop obtaining new information, that is, when the information collected is redundant \cite{ref36}). 

Since the explanatory model generated is mathematical in nature, an expert in mathematical modelling was also consulted, which allowed for the model to be finetuned.  

Once the information was validated by experts, it was collated to present a list of essential nutrients for the maintenance of the human body, classify these, and establish ideal intake levels. These data allowed us to establish a general model of nutritional utility. These results are presented and explained in more detail in the next section.

\section{Results}

\subsection{On nutrients: characteristics, classification, and adequate intake levels}

The documentary analysis and interviews with nutrition experts provided very consistent information on human nutritional needs. It should be noted that unlike in other areas of knowledge, we did not find any great debate in the literature on this topic or among the experts interviewed. On the contrary, the consensus was unequivocal. When we asked the interviewees about their opinions on the recommended intake values for essential nutrients, as presented in Table~\ref{tab1}, their responses showed consistent agreement. 

According to the literature, nutrients can be divided into macro- and micronutrients. Micronutrients are necessary in small amounts and need to be ingested through food or supplements because the body does not produce them by itself; these include vitamins, and minerals \cite{ref17}. Macronutrients should be consumed daily in large amounts since they constitute the source of energy required for human activities; these include proteins, carbohydrates, lipids, and water \cite{ref16, ref17, ref28}. In summary, the essential nutrients required by the body for normal functioning are vitamins, minerals, lipids, carbohydrates, fibre (which is a type of carbohydrate), proteins, and water.

As can be seen in Table~\ref{tab1}, nutritional requirements can vary significantly between men and women (Along with sex, experts point out that age also determines the required levels of nutrient intake. Thus, for example, the adequate intake of potassium is 4.7 grams per day for adults, but for infants (0–6 months) this is just 0.4 grams \cite{ref14}. Here we focus only on the adult population.).  Since in practice it is not possible to consume the exact listed value of each nutrient, the question arises: What are the repercussions of excess or insufficient intake? To answer this question, two limits must be considered in the development of a general model of nutrient intake: The \textit{Recommended Daily Allowance} (RDA) and the \textit{Tolerable Upper Intake Level} (UL).

The RDA specifies the daily amount of each nutrient the human body requires to be optimally healthy under normal conditions. Consuming less than the RDA implies an unsatisfied nutritional need, which can have consequences on one’s health. Sources such as Titchenal \cite{ref14} argue that for all nutrients, including water, having too much or too little has health consequences, e.g. affecting the levels of critical electrolytes in the blood. According to Titchenal \cite{ref14}, “RDA values are set to meet the [nutrient] needs of the vast majority (97 to 98 percent) of the target healthy population. […] The actual nutrient needs of a given individual will be different from the RDA.”

On the other hand, the documentation analyzed and the experts interviewed indicated that consuming more than these amounts (RDA) does not provide any additional health benefits, although it does not generally have negative effects either. However, there is a limit to harmless excess consumption, beyond which negative consequences can occur. Thus, most nutrients also have a UL, which is also presented in Table~\ref{tab1}. 

\begin{table} [H] 
\caption{Recommended daily allowance (RDA) and tolerable upper intake level (UL) of nutrients per day for adults.\label{tab1}}
\begin{tabularx}{1\textwidth}{X p{1.2cm} p{1.9cm} p{1.2cm} }
\toprule
\textbf{Nutrient}	& \textbf{RDA}	& \textbf{RDA} 	& \textbf{UL}\\
\textbf{Fat-soluble VITAMINS}	& \textbf{Men}	& \textbf{Women} 	& \textbf{ }\\
\midrule
Vitamin A (mcg)             & 900   & 700   & 3000 \\
Vitamin D (mcg)             & 15    & 15    & 50 \\
Vitamin K (mcg)             & 120   & 90    & - \\
Tocoferol/Vitamin E (mg)    & 15    & 15    & 1000 \\
\midrule
Water-soluble VITAMINS  \\		
\midrule
Vitamin C/Ascorbic acid (mg)& 90    & 75    & 2000\\
Vitamin B1/Thiamine (mg)    & 1.2   & 1.1   & -\\
Vitamin B2/Riboflavin (mg)  & 1.3   & 1.1   & -\\
Vitamin B3/Niacin (mg)      & 16    & 14    & 30\\
Pantothenic acid/Vitamin B5 (mg)& 5 & 5     & -\\
Vitamin B6 /Pyridoxine (mg) E3& 1.7 & 1.3   & 100\\
Folate/Folic acid/Vitamin B9(mcg)& 400 & 400 & 1000\\
Vitamin B12/Cyanocobalamin (mcg)& 2.4 & 2.4 & -\\
Biotin/Vitamin B7 \& B8 (mcg) & 30  & 30    & -\\
PROTEINS (g)                & 56    & 46    & -\\
CARBON HYDRATES (g)         & 325   & 225   & -\\
Dietary fibre (g)           & 38    &  25   & -\\
Water (varies by weight and age) & & & -\\
\midrule
MINERALS    	\\
\midrule
Calcium (mg)    & 	1000    & 1000    & 2500\\
Sodium (g)      & 	1.5     & 1.5     &	2.3\\
Potassium (g)	& 4.7       & 4.7     &	-\\
Phosphorus (mg)	& 700       & 700       & 4\\
Iron (mg)	    & 8         & 18        & 45\\
Iodine (mcg)	& 150       & 150       & 1100\\
Zinc (mg)	    & 11        & 8         & 40\\
Cobalt (mcg)	& 6         & 6         & -\\
Chromium (mcg)	& 35        & 25        & -\\
Manganese (mg)	& 2.3       & 1.8       & 11\\
Selenium (mcg)	& 55        & 55        & 400\\
Copper (mcg)	& 900       & 900       & 10000\\
Fluorine (mg)	& 4         & 3         & 10\\
Magnesium (mg)	& 400       & 310       & 350\\
Chlorine (g)	& 2.3       & 2.3       & 3.6\\
Sulfur	        & *\textsuperscript{1}  & *\textsuperscript{1}  & -\\
FATS (g)	    & 83        & 50        & -\\
Omega 3 (g)	    & 1.6       & 1.1       & -\\
Omega 6 (g)	    & 17        & 12        & -\\
\bottomrule
\end{tabularx}
\noindent{\footnotesize{\textsuperscript{1} * There is no recommended daily dose as the amounts required are very small.Sources: \cite{ref14, ref25, ref35, ref26, ref27}.
.}}
\end{table}
\unskip

\subsection{A general model of nutritional utility}

These nutrient characteristics, and others, are common knowledge among nutrition experts; however, these are not adequately reflected in the microeconomics models that study these needs. In the following, we will systematise the information collected in order to provide an adequate microeconomics modelling of it.

\begin{figure} 
\includegraphics[width=13.5 cm]{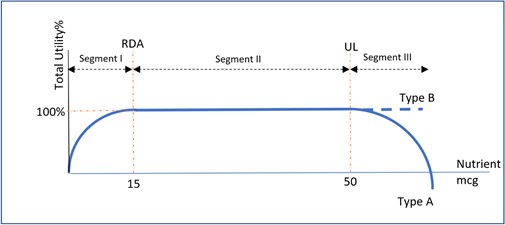}
\caption{Marginal utility from the intake of a nutrient (the case of vitamin D). Source: interviews and documentary analysis. Prepared by: the authors.\label{fig1}}
\end{figure}

\begin{figure} 
\includegraphics[width=13.5 cm]{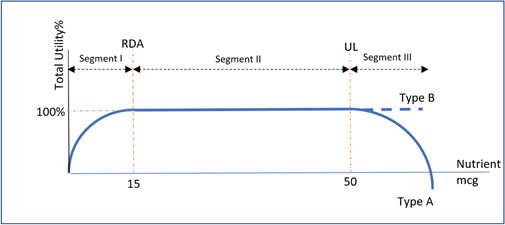}
\caption{Total utility (\%) for the intake of a nutrient (the case of vitamin D). Source: interviews and documentary analysis. Prepared by: the authors.\label{fig2}}
\end{figure}

The information gathered form the documentary research and the interviews shows that, in general, each nutrient has three different segments in its functional form of utility level based on the ingested amount in relation to the RDA and UL (see Figure~\ref{fig1} and Figure~\ref{fig2}).

\textit{Segment I}: intake is less than the RDA. In this case, the consumption level of said nutrient is inadequate, which can have a negative impact on health. For example, a low intake of iodine will affect thyroxine and triiodothyronine hormone production \cite{ref31}. Insufficient intake can negatively alter the processes of metabolism and affect brain development (e.g. brain damage), among other effects. According to Mikláš et al. \cite{ref32}, Zimmermann \cite{ref33}, and Lazarus \cite{ref34}, more than 2 billion people worldwide, which includes 30\% of people in European countries, are at risk of iodine deficiency.

There is no established functional relationship (specific equation) between the amount ingested at this first segment (I) and the benefits produced in the different parts of the human body (for example, between the amount of iodine ingested and the utility produced in the thyroid). However, from information gathered in this study, we can deduce that the relationship is positive monotonic (local), but not necessarily at a constant rate. That is, within segment I [0: RDA], the greater the quantity of the nutrient ingested, the better the health condition of the corresponding organs, e.g. iodine for the thyroid, vitamin A for the eyes, etc.

In any case, we can arbitrarily assign to this segment a behaviour that is similar to that determined by decreasing marginal utility.

\textit{Segment II}: intake is between the RDA and the UL. Any amount within this range brings no added benefits above segment I, but neither does it have negative consequences, i.e. the accumulated utility does not vary, which implies that the marginal utility in this range is zero. It should be noted that in some models, a \textit{saturation point} of maximum cumulative utility is proposed. This contrasts with the conclusions of this study, which also finds a \textit{saturation segment} but which, as we can verify with the data, is usually much larger in size than segment I.

\textit{Segment III}: intake is greater than the UL. In this context, two types of nutrients need to be differentiated: nutrients with defined UL (type A: w\_UL) and nutrients without UL (type B: wo\_UL). In the case of type A, the consumption of an amount greater than the UL generates a negative marginal utility that decreases the total accumulated utility; thus, for example, an excess of calcium above the UL can lead to kidney stones. In this case, a specific functional form cannot be established either, but we can assume negative monotonic behaviour. 

In the case of type B, we can assume that nutrients that do not have a defined UL have a marginal utility equal to zero (similar to segment II). As noted in Table~\ref{tab1}, some nutrients do not have a defined UL. This does not indicate that a UL does not exist, but rather that either the ordinary intake levels are so low that reaching the limit is unlikely (type A) or that the organism is perfectly capable of managing any excess (type B). We did not find a definitive explanation in the literature for the former, thus, in this case, the marginal utility remains zero and the total utility remains constant (slope of zero).

Figure~\ref{fig1} shows the marginal utility of the intake of a nutrient according to the three segments differentiated above. For example, in the case of vitamin D, segment II with zero marginal utility is more than twice as large as segment I; therefore, it cannot be excluded from a theoretical model.

Figure~\ref{fig2} plots the cumulative utility as a percentage of the nutritional need (RDA) covered by the intake of a nutrient across the three indicated segments.

All the experts stated that a nutrient intake above the RDA does not necessarily cause health problems as long as it does not exceed the UL. The interviewees agreed that in the case of nutrients for which a UL has been determined, excess intake above this limit causes health problems. More specifically, an intake of water-soluble vitamins greater than the UL does not have repercussions since the surplus is eliminated or disposed of from the body through the urine, but in the case of fat-soluble vitamins, there are negative consequences. Concerning the intake of minerals, proteins, fats, and carbohydrates, the experts agreed that consumption that exceeds the UL also causes damage to one’s health.

The experts interviewed agreed that one nutrient cannot be substituted for another in terms of function. However, three of the nine experts interviewed stated that in certain exceptional cases, the body can use proteins to perform the functions of fats, namely, to provide the body with energy. However, this substitution cannot occur in the opposite direction: the body cannot use fats to perform the functions of proteins.

\section{Discussion}

Nutrients are the material elements required for the body to function. They are ingested as integrated into natural and processed foods, but also in isolation as nutritional supplements. Each individual’s daily nutritional requirements vary depending on various factors such as age, sex, etc. However, the average requirements that are valid for most of the world's population have been established using scientific methods. Thus, the RDAs in Table~\ref{tab1} reflect the nutritional needs of between 97 to 98 percent of healthy adults \cite{ref14}. There is significant consensus regarding these data in the field of nutrition.

According to mainstream microeconomic theory, the marginal utility of a good decreases and can, at a certain point, become negative. This implies that the growth of accumulated total utility will slow down as it reaches its maximum point (saturation point) and, subsequently, will decline as a consequence of negative marginal utility. Thus, the model defines two segments to the left and right of the saturation point, which largely coincide with our proposed segments I and II.

\subsection{On nutritionally adequate intake} 

$ (RDA \leq intake \leq UL) $ 

An important difference to note is the saturation point of the classical model of indifference curves, which in our case is replaced by segment II. The latter, in addition, has the attribute of being significantly larger than segment I. For example, in the case of vitamin C the range of segment II is twenty times that of segment I.

When a particular nutrient is consumed in any point (amount) at segment II, its maximum utility is achieved. This is the same utility that would be achieved with the RDA since surplus below the UL (if existent) is harmless, with the nutrient in many cases being eliminated through urine, as in the case of vitamin C. This allows the body to properly manage the fact that the nutrient levels in food are not proportional to human requirements. Thus, for example, consuming a 250g serving of meat involves a 3\% excess intake (vitamin C) in relation to the respective RDA for men and 35\% more than the RDA for women, while for potassium this serving implies a 22\% excess consumption over the RDA for adults of both sexes. The situation is even more complex if we consider all of the nutrients present in that portion of meat.

It can be assumed that evolution itself has allowed the organism to develop the capacity to manage this situation by evacuating excesses nutrients, although only to a point. This homeostatic capacity to assimilate only the necessary amounts of nutrients is restricted by a limit (UL) above which there are negative repercussions on the organism. This limit is represented by segment II, where the functional form of utility can be represented by a line of slope zero since its marginal utility is zero.

\subsection{Characteristics of the functional form of utility generated by nutrient intake}

Based on the information collected, we can affirm that it is not possible to obtain a specific functional form for each nutrient $x_i$, much less a generalized one. However, some characteristics of these nutrient functional forms can be established and they are presented in Figure~\ref{fig1} and Figure~\ref{fig2}, and equation (1).

\begin{equation}
U(x_i) \% \left\{ \begin{array}{rclll} 
     <100\% & \mbox{for} & x_i<RDA_i              & \mbox{where} & du>0, d^2u<0 \\
     =100\% & \mbox{for} & RDA_i \le x_i \le UL_i & \mbox{where} & du=0, d^2u=0 \\
     <100\% & \mbox{for} & x_i>UL_i               & \mbox{where} & du<0, d^2u<0
     \end{array}\right. \\
\end{equation}

It should be emphasised that the RDAs are daily requirements (short-term), and many people have a \textit{permanent} (long-term) deficit or excessive intake, which is understood as an unhealthy diet and which will ultimately bring negative consequences.

\subsection{On the joint consumption of various nutrients}

We have carried out the analysis of the utility function generated by the intake of a nutrient $x_i$, which is equivalent to the satisfaction of an objective need for a specific RDA for each nutrient. Thus, if we ingest 100\% of each RDA indicated in Table~\ref{tab1}, then 100\% of our daily (physiological) nutritional needs will be satisfied. In this way, a daily utility function should satisfy the condition of equation (2):

\begin{equation}
U(x_1,x_2,\text{...},x_k )=100\%  \text{ if } x_i=RDA_i \text{ with i=1..k}.
\end{equation}

The daily utility obtained by a diet (adequate or not) that includes a bundle of nutrients can be modelled with a total daily utility function that is equivalent to the simple addition of the individual utility functions of each nutrient; although, in this case, the determination of the corresponding weights $\beta_i$ would remain pending.

\begin{equation}
U(x_1,x_2,\text{...},x_k )=\sum_{i=1}^{k} \beta_i u(x_i ) \text{ with i=1..k} \text{ and }\sum_{i=1}^{k} \beta_i =1
\end{equation}

If we ask which is more important, the heart, the brain, the liver, the immune system, etc., the most appropriate answer is that all of these parts are equally essential for life. “The eleven organ systems in the body completely depend on each other for continued survival as a complex organism” \cite{ref14}. Thus, we can justify the assignment of equal weights to every nutrient:

\begin{equation}
\beta_i=1/k \text{ for every }  i=1..k.
\end{equation}

The Cobb–Douglas utility function is very frequently used in the literature; however, in this context it does not adequately capture the characteristics that correspond to the objective information collected. Thus, for example, with this function being a product the total utility obtained in a day is zero if in that period a particular nutrient is not ingested, even if the RDA of all others is consumed. This result does not correspond to reality. On the other hand, the Cobb–Douglas function cannot adequately represent and combine the mixed structure of the individual utility function in equation (1).

In the case of complementary goods, the Leontief utility function is frequently proposed. In this case, we also have the problem that the null intake of a nutrient on a specific day would imply that the total utility is zero (minimum). On the other hand, there is an RDA for each nutrient but its intake is not conditioned to be proportional in each minor (deficient) situation. For example, if a diet is deficient (minimal) in only one nutrient, the RDAs of the others may still be reached. In fact, in the real world, it is very common for a deficiency of a particular nutrient to be diagnosed without a proportional deficiency in the others.

\subsection{On indifference curves and the indifference area}

\begin{figure}[H]
\includegraphics[width=13.5 cm]{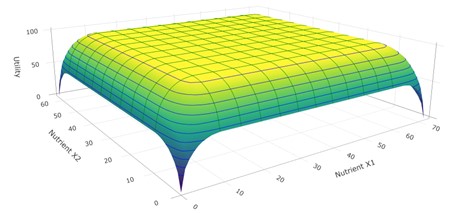}
\caption{Total utility from the intake of two nutrients $x_1$ and $x_2$, type A. Prepared by: the authors.\label{fig3}}
\end{figure}

Based on the above, if we assume that the nutrient bundle includes only two nutrients, $x_1$ and $x_2$, then in terms of human physiology the utility function can be graphically represented as in Figure~\ref{fig3}.

In Figure~\ref{fig3}, the horizontal axes represent the ingested amounts of the nutrients and the vertical axis corresponds to the total utility obtained by the combination of different intake levels for these. If none of nutrient $x_1$ is ingested, the utility generated corresponds only to the utility function of nutrient $x_2$, and vice versa. Obviously, in areas with an increasing or decreasing slope, we find contour curves that correspond to the criteria of indifference curves, that is, to those combinations of nutrients that generate the same level of total utility. However, in this case, the horizontal area stands out (with slope zero), and this corresponds to the combination of segment II intake for each nutrient. In this case, there is an \textit{indifference area} of significant magnitude that cannot be ignored in the modelling of the phenomenon. This area contradicts the assumption of local non-satiation disacussed in microeconomics textbooks such as Mas-Colell et al. \cite{ref20} and Varian \cite{ref24}, in which it is argued that this assumption (local non-satiation) “rules out ‘thick’ indifference curves”.

In this indifference area, the assumption of local non-satiation is not admissible, and neither is increasing or decreasing monotonic behaviour. It is an area in which all combinations of nutrients provide the same total utility: the maximum possible (100\%). Obviously, an area of indifference also implies that all indifference curves traceable to it can cross each other, which was apparently not possible according to traditional microeconomic theory.

If in this analysis we include type B nutrients (without a UL), then the area of indifference is even larger since a decreasing utility is not included in segment III.

\subsection{On the impossibility of nutrient substitution}

The total utility function discussed above implies that the utility generated by nutrient $x_1$ can be added to the utility generated by nutrient $x_2$ according to equation  (5).

\begin{equation}
U(x_1,x_2 )=\beta_1 u(x_1 )+\beta_2 u(x_2)
\end{equation}

This total utility function, as well as the others mentioned, assumes that the unit in which utility is measured is common for the two nutrients combined and that there is a substitution rate between them. These two characteristics are related, as we will see below.

For example, about 200 milligrams of sodium are required daily. This mineral is essential for maintaining fluids in a state of balance \cite{ref14}—this is its utility (among others). Thus, if less than that RDA is ingested, fluid balance will not be properly regulated. Thus, this mineral has the specific utility of preserving the balance of fluids. On the other hand, vitamin A is needed to make the pigment rhodopsin, which \textit{allows eyes to see under low-light conditions} \cite{ref14}, that is, it has the specific utility of allowing eyes to see under low-light conditions. If we consider the satisfaction of specific utilities, which require the intake of the RDA of a specific nutrient, there is no amount of sodium that generates the utility of allowing us to see under low-light conditions, nor is there an amount of vitamin A that generates the utility of preserving fluid balance. At this level of need, it is not possible to substitute one nutrient for another since the specific utilities are different and, therefore, non-addable.

However, we can consider the need from other, more abstract levels. Thus, the intake of vitamin A helps maintain healthy sight, but it also collaborates, in an abstract way, with an individual’s well-being and, in a more diffuse (abstract) way, with family and even societal well-being. At this abstract level, concrete needs can be seen as important components of a greater complex whole that includes subjective values. Synthetic indicators can be constructed for the measurement of this subjective complexity, which then presents the problem of determining the assigned weights for each nutrient, as well as determining the functional form of aggregation (These are important aspects but are outside the scope of this study).

In this study, we focus solely on the specific level of nutritional need, which, based on the information collected, corresponds to the reality in which each nutrient fulfils a physiological role and is not interchangeable with another—an objective characteristic that prevents establishing substitution rates. Thus, when we analyse the possible combinations of sodium ($x_1$) and vitamin A ($x_2$) and determine their contribution to the concrete utility $U_{fb} (x_1,x_2 )$ of satisfying the need to preserve fluid balance, then we must specify that

\begin{equation}
U_{fb} (x_1,x_2 )=\beta_1 u_{fb} (x_1 )+\beta_2 u_{fb} (x_2 )
\end{equation}

And with  $u_{fb} (x_2 )=0$ for any value of $x_2$, then

\begin{equation}
U_{fb} (x_1,x_2 )=u_{fb} (x_1 ).
\end{equation}

\begin{figure}[H]
\includegraphics[width=13.5 cm]{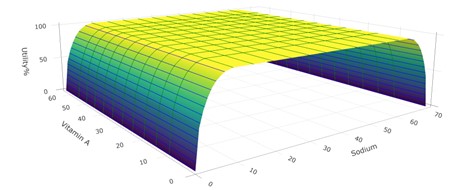}
\caption{Concrete total utility of ‘preserving fluid balance’ through sodium and vitamin A intake. Note: No amount of vitamin A provides a concrete utility for ‘preserving fluid balance’. Prepared by: the authors.\label{fig4}}
\end{figure}

This can be represented as in Figure~\ref{fig4}.

It is noteworthy that, in this case, a significantly large indifference area is also generated in the model. On the other hand, indifference curves are clearly defined as indifference lines. These indifference lines are lines perpendicular to the sodium axis in  Figure~\ref{fig4}.

\section{Conclusion} 

The main objective of this study was to build a theoretical model that adequately characterizes the utility obtained by satisfying human nutritional needs, and compate this to the conventional model of indifference curves.

Based on the information collected, we have established that a nutritional utility model must include the presence of indifference areas in which "more is not better or worse". These are areas in which, in contrast to the principles of conventional theory, an infinity of curves that intersect can be drawn because they all correspond to the maximum utility surface. Also relevant is the fact that each nutrient has a specific utility, or metabolic function, in the organism that, in general, cannot be provided by another nutrient. This implies that it is not possible to establish a substitution rate (As can be seen in Figure 4, the slopes of the indifference lines are infinite in segments I and III and could not be determined in segment II since there is no indifference curve but, rather, an indifference area). 

It should be reiterated that in today's society it is very common to opt for a diet based primarily on objective nutritional needs in the case of the diets of children, the elderly, the sick, and even healthy adults, but that many choose to consume food solely according to their gastronomic preferences (subjective tastes), which may clash with established nutritional (objective) needs. When the \textit{desire} does not coincide with the nutritional \textit{need}, this divergence can lead to significant health consequences, such as childhood obesity, and is therefore an important field of study. However, there are the effects of poverty, such as structural inequalities in society and the unaffordable cost of healthy eating in many places, a major reason why many people have poor eating habits.

Governments and many institutions such as the World Health Organization (WHO) are primarily concerned with promoting a diet based on objective nutritional needs. It can be assumed that the mismatch between desire and physiological need may imply future problems, at least in the field of human nutrition. In many cases this can lead to a situation of regret for one’s choices, a phenomenon that can be addressed from the perspective of regret theory \cite{ref20}.

Finally, although the conclusions of this research are limited to the field of human nutrition, future studies should carry out similar work at the other levels of Maslow’s hierarchy, since it is pertinent to question to what extent a theoretical rate of substitution of food for cell phones makes sense, for example, since in practical terms no amount of cell phones can cover the need for food, and vice versa.

\bibliographystyle{ieee}


\end{document}